\documentclass[preprint,aps,nofootinbib]{revtex4}

\usepackage{graphicx}

\begin{document}

\title{Thermal leptogenesis with triplet Higgs boson and mass varying neutrinos}
\author{Peihong Gu}
\email{guph@mail.ihep.ac.cn}
\author{Xiao-Jun Bi}
\affiliation{Institute of High
Energy Physics, Chinese Academy of Sciences, P.O. Box 918-4,
Beijing 100039, People's Republic of China}

\begin{abstract}
We study the thermal leptogenesis in the scenario where the
standard model is extended to include one $SU(2)_{L}$ triplet
Higgs boson, in addition to three generations of the right-handed
neutrinos. And in the model we introduce the coupling between the
Quintessence and the right-handed neutrinos, the triplet Higgs
boson, so that the light neutrino masses vary during the evolution
of the universe. Assuming that the lepton number asymmetry is
generated by the decays of the lightest right-handed neutrino
$N_{1}$, we find the thermal leptogenesis can be characterized by
four model independent parameters. In the case where the
contribution of the triplet Higgs to the lepton asymmetry is
dominant, we give the relation between the minimal $M_{1}$ and the
absolute mass scale $\bar{m}$ of the light neutrinos, by solving
the Boltzmann equations numerically. We will also show that with
the varying neutrino masses, the reheating temperature can be
lowered in comparison with the traditional thermal leptogenesis.
\end{abstract}

\maketitle

The baryon number asymmetry of the Universe has been determined
precisely\cite{max}:
\begin{equation}
\eta_B\equiv \frac{n_B}{n_\gamma}=(6.3 \pm 0.3)\times 10^{-10}\ ,
\end{equation}
where $n_B=n_b-n_{\bar{b}}$ and $n_\gamma$ are the baryon and
photon number densities, respectively. As the confirmation of the
neutrino oscillations by several experiments\cite{ahn,ahm},
leptogenesis\cite{lepto} is now an attractive scenario to explain
the observed baryon number asymmetry, where the lepton number
asymmetry is first produced and then converted to the baryon
number asymmetry via the $(B+L)$-violating sphaleron
interactions\cite{spha}.

The minimal thermal leptogenesis is quite economic and only
requires three generations of the right-handed Majorana neutrinos
beyond the standard model, which are also necessary to explain the
small neutrino masses through the seesaw mechanism\cite{seesaw}.
However, this scenario seems require too high reheating
temperature which may conflict with the upper bound of the
reheating temperature set by the gravitino problem\cite{grav}, and
hierarchical neutrino spectrum with $m_i\lesssim 0.12
eV$\cite{buch1}. Furthermore, if the light neutrinos are
degenerate as indicated by the experimental signal of neutrinoless
double beta decay\cite{0neu}, it is hard to imagine that the Dirac
and the Majorana neutrino mass matrices, both naturally having
hierarchical eigenvalues, conspire to produce the degenerate
neutrino spectrum via the seesaw mechanism. The degenerate
neutrino spectrum is naturally produced in the type II
seesaw\cite{shafi} model, where a triplet Higgs boson is
introduced, whose vacuum expectation value ($vev$) gives the
common neutrino mass scale and the type I seesaw produces the mass
square differences required by the oscillation experiments.

A way to reconcile the minimal thermal leptogenesis and the
gravitino problem is to consider that the neutrino masses are
cosmological variable\cite{gu,nelson}. In a recent work we studied
the scenario\cite{bi} where the interaction between the
right-handed neutrinos and the
Quintessence\cite{ratra,wett,frie,zlatev}, a dynamical scalar
field as a candidate for the dark energy\cite{turner}, which
drives the accelerating of the Universe at the present
time\cite{pel}, makes the masses of the right-handed neutrinos
vary during the evolution of the universe. In this scenario, the
reheating temperature is lowered and compatible with the limits
set by the gravitino problem, and degenerate light neutrino
spectrum is also permitted. However, the two conditions, the low
reheating temperature and degenerate neutrino spectrum, can not be
satisfied simultaneously yet.

In this paper, we study a scenario of the thermal leptogenesis
where the light neutrinos are degenerate, to explain the
neutrinoless double beta decay, and at the same time the reheating
temperature is low. This scenario is realized in the type II
seesaw model with variable neutrino masses. The type II seesaw
model including one $SU(2)_{L}$ triplet Higgs boson\cite{sen}, in
addition to three generations of the right-handed neutrinos, is a
general scenario derived from the left-right symmetric
models\cite{laz}.

In this scenario, the lepton number asymmetry can be generated by
the decays of the right-handed neutrinos and/or the $SU(2)_{L}$
triplet Higgs\cite{zhang}. Assuming a hierarchical right-handed
neutrino spectrum, $M_{1}$ $\ll$ $M_{2},M_{3}$, and $M_1$ is also
much lighter than the mass of the triplet Higgs $M_1\ll
M_{\Delta}$, the lepton number asymmetry comes mainly from the
decays of the lightest right-handed neutrino $N_{1}$. We find the
thermal leptogenesis can be characterized by four model
independent parameters: the $CP$ asymmetry $\varepsilon_{1}$ of
$N_{1}$ decays, the heavy neutrino mass $M_{1}$, the absolute mass
scale $\bar{m}$ of the light neutrinos, and the effective light
neutrino mass $\tilde{m}_{1}$, which is a similar result as in the
minimal thermal leptogenesis\cite{buch2}.

The Lagrangian relevant to leptogenesis
reads:
\begin{equation}
-\mathcal{L}=\frac{1}{2}M_{i}\bar{N}^{C}_{Ri}N_{Ri}
+M_{\Delta}^{2}Tr\Delta_{L}^{\dagger}\Delta_{L}
+g^{\nu}_{ij}\bar{\psi}_{Li}N_{Rj}\phi
+g^{\Delta}_{ij}\bar{\psi}^{C}_{Li}i\tau_{2}\Delta_{L}\psi_{Lj}
-\mu\phi^{T}i\tau_{2}\Delta_{L}\phi +h.c.
\end{equation}
where $\psi_L=(\nu, l)^{T}$, $\phi=(\phi^{0}, \phi^{-})^{T}$ are
the lepton and the Higgs doublets, and
\begin{displaymath}
\Delta_{L}=  \left(\begin{array}{cc}
\frac{1}{\sqrt{2}}\delta^{\dagger} & \delta^{\dagger\dagger} \\
\delta^{0} & -\frac{1}{\sqrt{2}}\delta^{\dagger}
\end{array}\right)
\end{displaymath}
is the Higgs triplet.

After the electroweak phase transition, the left-handed neutrino
mass matrix can be written as
\begin{equation}
\label{mneu}
m_{\nu}=-g^{\nu*}\frac{1}{M}g^{\nu\dagger}v^{2}+2g^{\Delta}v_{L}=m_{\nu}^{I}+m_{\nu}^{II}
\end{equation}
where $m_{\nu}^{I}$ is the type I seesaw mass term, $m_{\nu}^{II}$
is the type II seesaw mass term, and $v=174GeV$,
$v_{L}\simeq\mu^{*}v^{2}/M_{\Delta}^{2}$ are the $vev$s of $\phi$
and $\Delta_{L}$, respectively.

\begin{figure}
\includegraphics[scale=.8]{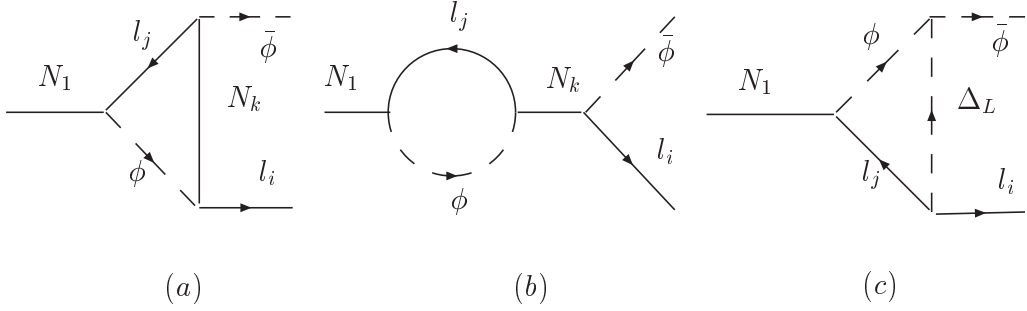}
\caption{The loop-diagrams of $N_{1}$ decays.}
\end{figure}

The $CP$ asymmetry $\varepsilon_{1}$ is generated by the
interference of the loop diagrams, shown in Fig. 1, with the tree
diagram of $N_1$ decay. Besides the two same diagrams as in the
minimal seesaw scenario, diagrams (a) and (b) in Fig. 1, there is
an additional diagram (c) with the exchange of the Higgs triplet.
We then have
\begin{equation}
\varepsilon_{1}=\varepsilon_{1}^{N}+\varepsilon_{1}^{\Delta}\ ,
\end{equation}
with $\varepsilon_{1}^{N}$ and $\varepsilon_{1}^{\Delta}$ the $CP$
asymmetry of $N_{1}$ decays due to the exchange of the
right-handed neutrinos and the Higgs triplet, respectively. For$
M_{1}\ll M_{2}, M_{3}, M_{\Delta}$, we have\cite{covi,king}
\begin{equation}
\varepsilon_{1}^{N}\simeq \frac{3}{16 \pi}\frac{M_{1}}{v^{2}}\frac
{\sum_{ij}\text{Im}[g^{\nu\dagger}_{1i}g^{\nu\dagger}_{1j}(m^{I*}_{\nu
})_{ij}]}{(g^{\nu\dagger}g^{\nu})_{11}}\ ,
\end{equation}
\begin{equation}
\varepsilon_{1}^{\Delta} \simeq \frac{3}{16 \pi}
\frac{M_{1}}{v^{2}} \frac{\sum_{i,j} \text{Im}[g^{\nu
\dagger}_{1i}g^{\nu \dagger}_{1j} (m^{II*}_{\nu})_{ij}]}{(g^{\nu
\dagger}g^{\nu})_{11}}\ .
\end{equation}
It is interesting to see that the triplet contribution will
dominate the $CP$ asymmetry $\varepsilon_{1}$ when it dominates
the neutrino mass matrix $m_{\nu}$\cite{sen}. In this case, there
is an upper bound on the asymmetry\cite{sen,king,ib,buch2},
\begin{equation}
\mid \varepsilon_{1}\mid \simeq \mid \varepsilon_{1}^{\Delta}\mid
\leq \frac{3M_{1}m_{3}}{16 \pi v^{2}} \simeq
\varepsilon_{1}^{max}\ .
\end{equation}

In order to calculate the final baryon number asymmetry, we have
calculated the washout effect by solving the Boltzmann equations
numerically. All the relevant processes should be taken into
account, which include $N_{1}$ decays and inverse-decays; the
$\Delta L=1$ scatterings mediated by exchanging doublet Higgs; and
$\Delta L=2$ scatterings mediated by exchanging the right-handed
neutrinos and the triplet Higgs. By solving the Boltzmann
equations, we can get the baryon-to photon ratio $\eta_{B}$.

In comparison with the minimal seesaw
scenario\cite{luty,plum,buch2,giudice}, there new $\Delta L=2$
scattering processes with the exchanges of the triplet Higgs
should be considered. We give the reduced cross sections
$\hat{\sigma}_{N}$ for the process $l\bar{\phi} \leftrightarrow
\bar{l}\phi$, and $\hat{\sigma}_{N,t}$ for the process
$ll(\bar{l}\bar{l}) \leftrightarrow
\phi\phi(\bar{\phi}\bar{\phi})$ as:
\begin{displaymath}
\hat{\sigma}_{N(N,t)}(s)=\frac{1}{2
\pi}[\sum_{i}(g^{\nu\dagger}g^{\nu})_{ii}^{2}f^{N(N,t)}_{ii}(x)+
\sum_{i,j}Re[(g^{\nu\dagger}g^{\nu})_{ij}^{2}]f^{N(N,t)}_{ij}(x)
\end{displaymath}
\begin{equation}
+ \sum_{i,j,k}Re
[g^{\Delta\dagger}_{ij}g^{\nu\dagger}_{ki}g^{\nu\dagger}_{kj}\frac{\mu}{M_{1}}]f^{\Delta
N(\Delta N,t)}_{ijk}(x)+
\sum_{i}(g^{\Delta\dagger}g^{\Delta})_{ii}\frac{
\mid\mu\mid^{2}}{M_{1}^{2}}f^{\Delta(\Delta ,t)}_{ii}(x)]\ ,
\end{equation}
with
\begin{equation}
f^{N}_{ii}(x)=1+\frac{a_{i}}{D_{i}(x)}+\frac{xa_{i}}{2D_{i}^{2}(x)}
-\frac{a_{i}}{x}[1+\frac{x+a_{i}}{D_{i}(x)}]\log(1+\frac{x}{a_{i}})\
,
\end{equation}
\begin{displaymath}
f^{N}_{ij}(x)=\frac{1}{2}\sqrt{a_{i}a_{j}}[\frac{1}{D_{i}(x)}+\frac{1}{D_{j}(x)}+\frac{x}{D_{i}(x)D_{j}(x)}
\end{displaymath}
\begin{equation}
+(1+\frac{a_{i}}{x})(\frac{2}{a_{j}-a_{i}}-\frac{1}{D_{j}(x)})\log(1+\frac{x}{a_{i}})
+(1+\frac{a_{j}}{x})(\frac{2}{a_{i}-a_{j}}-\frac{1}{D_{i}(x)})\log(1+\frac{x}{a_{j}})]\
,
\end{equation}
\begin{equation}
f^{\Delta}_{ii}(x)=12[\frac{1}{x}\log(1+\frac{x}{y})-\frac{1}{x+y}]\
,
\end{equation}
\begin{equation}
f^{\Delta N
}_{ijk}(x)=8\frac{\sqrt{a_{k}}(x-a_{k})}{(x-a_{k})^{2}+a_{k}c_{k}}
[1-\frac{y}{x}\log(1+\frac{x}{y})]+4\frac{\sqrt{a_{k}}}{x+a_{k}+y}
[\frac{y}{x}\log(1+\frac{x}{y})-(1+\frac{a_{k}}{x})\log(1+\frac{x}{a_{k}})]\
,
\end{equation}
\begin{equation}
f^{N,t}_{ii}(x)=\frac{x}{x+a_{i}}+\frac{a_{i}}{x+2a_{i}}\log(1+\frac{x}{a_{i}})\
,
\end{equation}
\begin{equation}
f^{N,t}_{ij}(x)=\frac{1}{2}\frac{\sqrt{a_{i}a_{j}}}{(a_{i}-a_{j})(x+a_{i}+a_{j})}
[(2x+3a_{i}+a_{j})\log(1+\frac{x}{a_{j}})-(2x+3a_{j}+a_{i})\log(1+\frac{x}{a_{i}})]\
,
\end{equation}
\begin{equation}
f^{\Delta,t}_{ii}(x)=6\frac{x(x-y)^{2}}{[(x-y)^{2}+yc_{\Delta}]^{2}}\
,
\end{equation}
\begin{equation}
f^{\Delta N,t}_{ijk}(x)=
6\sqrt{a_{k}}\frac{x-y}{(x-y)^{2}+yc_{\Delta}}\log(1+\frac{x}{a_{k}})\
.
\end{equation}
Here $x=\frac{s}{M_{1}^{2}}$, $a_{i}\equiv M_{i}^{2}/M_{1}^{2}$
and $1/D_{i}(x)\equiv (x-a_{i})/[(x-a_{i})^{2}+a_{i}c_{i}]$ is the
off-shell part of the $N_{i}$ propagator with $c_{i}\equiv
a_{i}(g^{\nu\dagger}g^{\nu})^{2}_{ii}/(8 \pi)^{2}$, and
$c_{\Delta}\equiv \Gamma _{\Delta}^{2}/M_{1}^{2}$. Similar to Ref.
\cite{buch2}, the reaction density $\gamma_{N}+\gamma_{N,t}$ can
be separated into two parts: the resonance contribution which is
highly peaked around $x=1$ and the contribution comes from the
region $x\ll1$ which corresponds to $z\gg1$,
\begin{equation}
\gamma_{N}^{res}=\frac{M_{1}^{5}}{64
\pi^{3}v^{2}}\tilde{m}_{1}\frac{1}{z}K_{1}(z)\ ,
\end{equation}
\begin{equation}
\gamma_{N}(z\gg1)\simeq\gamma_{N,t}(z\gg1)\simeq
\frac{3M_{1}^{6}}{8 \pi^{5}v^{4}}\bar{m}^{2}\frac{1}{z^{6}}\ ,
\end{equation}
where $\tilde{m}_{1}\equiv
\frac{(g^{\nu\dagger}g^{\nu})_{11}v^{2}}{M_{1}}$ and
$\bar{m}^{2}\equiv
m_{1}^{2}+m_{2}^{2}+m_{3}^{2}=\text{tr}(m_{\nu}^{\dagger}m_{\nu})
=\text{tr}((m_{\nu}^{I}+m_{\nu}^{II})^{\dagger}(m_{\nu}^{I}+m_{\nu}^{II}))$.
Analysis for $z<1$ is also similar to Ref.\cite{buch2}. The
reaction densities $\gamma_{i}$ \cite{luty} is defined as:
\begin{equation}
\gamma_{i}(z)=\frac{M_{1}^{4}}{64 \pi^{4}}\frac{1}{z}
\int^{\infty}_{(m_{a}^{2}+m_{b}^{2})/M_{1}^{2}}dx\hat{\sigma}_{i}(x)\sqrt{x}K_{1}(z\sqrt{x})\
,
\end{equation}
where $m_{a}$ and $m_{b}$ are the masses of the two particles in
the initial state. Since $\gamma_{N}+\gamma_{N,t}$ is not changed
compared with the results of Ref.\cite{buch2}, the thermal
leptogenesis can be still characterized by four parameters:
$\varepsilon_{1}$, $M_{1}$, $\bar{m}$, $\tilde{m}_{1}$, even in
the presence of the $SU(2)_{L}$ triplet Higgs.

The washout effect mainly depends on the effective neutrino mass
$\tilde{m}_{1}$. Since $m_{1} \leq \tilde{m}_{1} \lesssim
m_{3}$\cite{ib,buch2} should not be satisfied when the triplet
contribution is dominant, we can always adjust $ \tilde{m}_{1}$ to
avoid the large washout effect. Therefore the neutrino mass
spectrum can be degenerate, even above the cosmological
bound\cite{sen,chun,rode}. The numerical result is shown in Fig.
2. For $\bar{m}\simeq 0.051eV$, the low limit of the neutrino mass
scale from the oscillation experiments
constraint\cite{ahn,ahm,buch3}, we get $M_{1}\simeq 3.4 \times
10^{9}GeV$. We can get $M_{1}\simeq 2.7\times 10^{8}GeV$ for
$\bar{m}\simeq 1.0eV$, which is the upper bound from the
cosmological constraint $ \sum_{i} m_{i} < 1.7eV$\cite{max}.

We notice that these values of $M_{1}$ are only marginally
consistent with the bound set by the gravitino problem\cite{grav}.
In order to solve the problem, we consider the light neutrino
masses are varying during the evolution of the universe\cite{gu}.
We introduce a parameter $k$ which indicates the ratio of the
light neutrino masses at the leptogenesis epoch and the present
epoch. When solving the Boltzmann equations, the $M_{1}, \bar{m}$,
and $\tilde{m}_{1}$ should all take the values at the leptogenesis
epoch. If $\bar{m}$ takes the value at the present epoch, we
should replace $\bar{m}$ by $k\bar{m}$ in the Boltzmann equations.
By solving the Boltzmann equations numerically, we can see the
reheating temperature are lowered with the increasing $k$. For
$k=10$, we can get $M_{1} \simeq 3.1 \times 10^{8}GeV$ for
$\bar{m}\simeq 0.051eV$, and $M_{1} \simeq 2.7 \times 10^{7}GeV$
for $\bar{m} \simeq 1.0eV$. $M_{1}$ is lowered to $M_{1} \simeq
3.1 \times 10^{7}GeV$ with $\bar{m}\simeq 0.051eV$ and $M_{1}
\simeq 2.5 \times 10^{6}GeV$ with $\bar{m}\simeq 1.0eV$ for
$k=100$. In this paper, the vales of $M_{1}$ are all at the
leptogenesis epoch, and $\bar{m}$ takes the value at the present
epoch.
\begin{figure}
\includegraphics[scale=1.5]{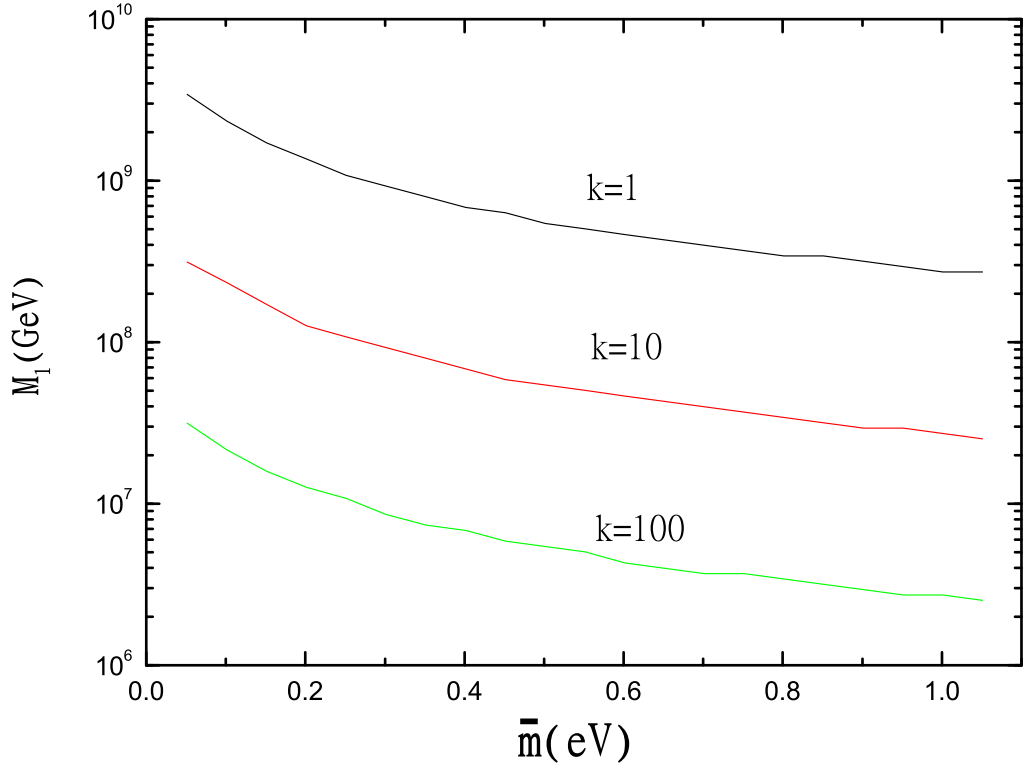}
\caption{The minimal $M_{1}$ as a function of $\bar{m}$ for
$\eta_{B}=5.4 \times 10^{-10}$. $M_{1}$ is the value of the
right-handed neutrino mass at the leptogenesis epoch, and
$\bar{m}$ is the absolute mass scale of the light neutrino at the
present epoch. The curve with $k=1$ stand for the case without the
variation of the light neutrino masses.}
\end{figure}

In this paper, we get the varying neutrino masses by introducing
the interaction between the Quintessence and the right-handed
neutrinos, the triplet Higgs in Eq.$(2)$. Assume these
interactions take simple forms as:
\begin{equation}
M_i \rightarrow M_i(Q)=\bar{M}_i e^{\beta \frac{Q}{M_{pl}}}\ ,
\end{equation}
\begin{equation}
M_\Delta \rightarrow M_\Delta(Q)=\bar{M}_\Delta e^{\frac{\beta}{2}
\frac{Q}{M_{pl}}}\ ,
\end{equation}
where $\beta$ is a $\mathcal{O}(1)$ coefficient. Then we get
\begin{equation}
m_{\nu}\propto e^{-\beta \frac{Q}{M_{pl}}}\ ,
\end{equation}
and
\begin{equation}
k= e^{\beta \frac{Q_{0}-Q_{D}}{M_{pl}}}\ .
\end{equation}
$Q_{0}$, $Q_{D}$ are the values of the Quintessence field at the
present epoch and the leptogenesis epoch, respectively.

For a numerical calculation of $k$, we consider a model of the
Quintessence with the double exponential potential\cite{Barreiro}
\begin{equation}
V=V_0(e^{\lambda Q}+e^{\alpha Q})\ .
\end{equation}
The equations of motion of the Quintessence, which for a flat
universe, are given by,
\begin{eqnarray}
&&H^2 =\frac{8 \pi G}{3} (\rho_B +\frac{\dot{Q}^{2}}{2}+V(Q)),\;\\
&&\ddot{Q}+3H\dot{Q}+\frac{dV(Q)}{dQ}=0\ ,
\end{eqnarray}
where $\rho_B$ represent the energy densities of the background
fluid. This model has the tracking property for suitable
parameters. Here, we choose $\lambda =100M_{pl}^{-1}$, $\alpha
=-100M_{pl}^{-1}$, the initial value of Quintessence field
$Q_i=1.374 M_{pl}$ and for the equation-of-state, which is defined
as
\begin{equation}
\label{wq} w_Q=\frac{{\dot{Q}^2}/{2}-V(Q)} {{\dot{Q}^2}/{2}+V(Q)}\
\ ,
\end{equation}
the initial value is $w_{Q_i}=-1$. We obtain that $\Omega_{Q_0}
\simeq 0.72$ and the present equation-of-state of the Quintessence
is $w_{Q_0}\simeq -1$ which are consistent with the observational
data. In Fig. 3, we show the evolution of $w_Q$ and $Q$ with the
temperature $T$.

\begin{figure}
\includegraphics[scale=1.5]{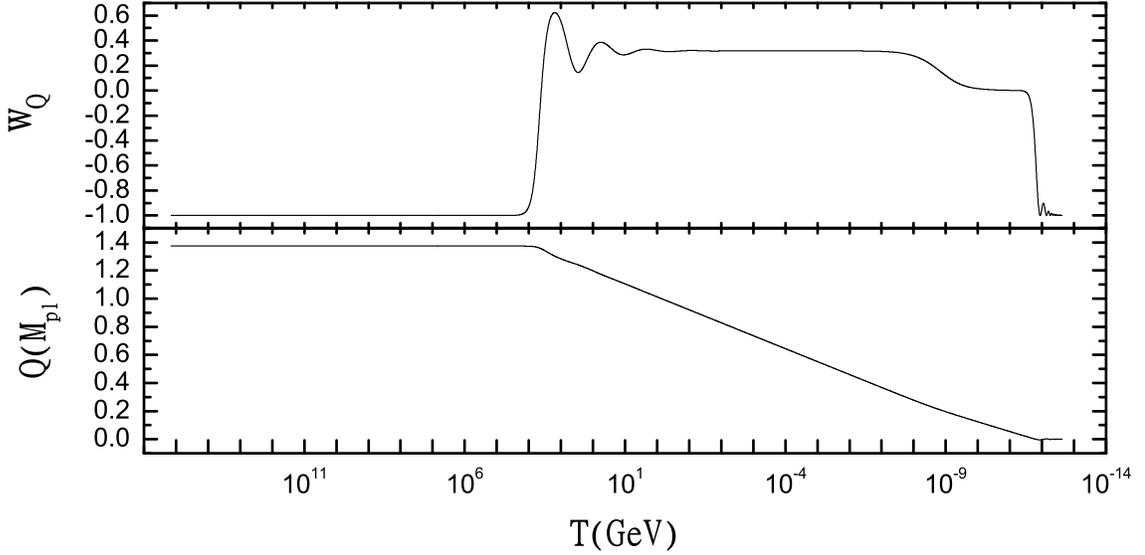}
\caption{The evolution of $w_Q$ and $Q$ as a function of the
temperature $T$ for the double exponential Quintessence model.}
\end{figure}

Taking into account the interaction with the right-handed
neutrinos and the triplet Higgs, we get the equation of motion of
the Quintessence as
\begin{equation}
\label{vi}
\ddot{Q}+3H\dot{Q}+\frac{dV(Q)}{dQ}+\frac{dV_I(Q)}{dQ}=0\ .
\end{equation}
The source term in the equation above is given by\cite{peebles}
\begin{displaymath}
\frac{dV_I(Q)}{dQ}=\sum_in_i\frac{dM_i}{dQ}
\left\langle\frac{M_i}{E_{i}}\right\rangle+n_\Delta\frac{dM_\Delta}{dQ}
\left\langle\frac{M_\Delta}{E_{\Delta}}\right\rangle
\end{displaymath}
\begin{equation}
=\frac{\beta}{M_{pl}}\frac{1}{\pi^2}T\sum_i
M_i^3K_1(M_i/T)+\frac{3}{4}\frac{\beta}{M_{pl}}\frac{1}{\pi^2}T
M_\Delta^3K_1(M_\Delta/T)\ ,
\end{equation}
where $n_{i}$ and $E_{i}$ are the number density and the energy of
the right-handed neutrinos respectively, $n_\Delta$ and
$E_{\Delta}$ belong to the triplet Higgs, $\langle \rangle$
indicates thermal average, and $K_1$ is the modified Bessel
function. For simplicity, we have taken the Maxwell-Boltzmann
distribution of the right-handed neutrinos and the triplet Higgs
in the last step of the equation.

We then solve the equation (\ref{vi}) numerically, assuming
$\overline{M}_3=10\overline{M}_2=10^{4}\overline{M}_1$ and
$\overline{M}_\Delta=10^{3} \overline{M}_1$. The numerical results
are shown in Figs. 4, 5, 6, and 7, where we have taken the same
definition of $w_Q$ as in Eq.(27). In Fig. 4 and 5, we take
$\beta=-1.68$, $\overline{M}_1 =3.1 \times 10^9$ GeV, and
$\beta=-3.35$, $\overline{M}_1 =3.1 \times 10^9$ GeV,
respectively, which give rise to $Q_0 \simeq 0$ and $Q_D \simeq
1.374 M_{pl}$. We then have $M_1 \simeq 3.1\times 10^8 GeV$ for
$k=10$, and $M_1 \simeq 3.1\times 10^7 GeV$ for $k=100$,
corresponding to the case we considered in the hierarchical
neutrino spectrum with $\bar{m}\simeq 0.051eV$. In Fig. 6 and 7,
we choose the parameters $\beta=-1.68$, $\overline{M}_1 =2.7
\times 10^{8} GeV$
 and $\beta=-3.35$, $\overline{M}_1 =2.5 \times 10^{8} GeV$, respectively. We find
the values of $Q_0$ and $Q_D$ are almost the same as the above
case. We then have $M_1 \simeq 2.7\times 10^{7} GeV$ for $k=10$
and $M_1 \simeq 2.5\times 10^{6} GeV$ for $k=100$, corresponding
to the case that satisfies the cosmic limit $\bar{m}\simeq 1.0eV$
for the degenerate neutrinos.
\begin{figure}
\includegraphics[scale=1.5]{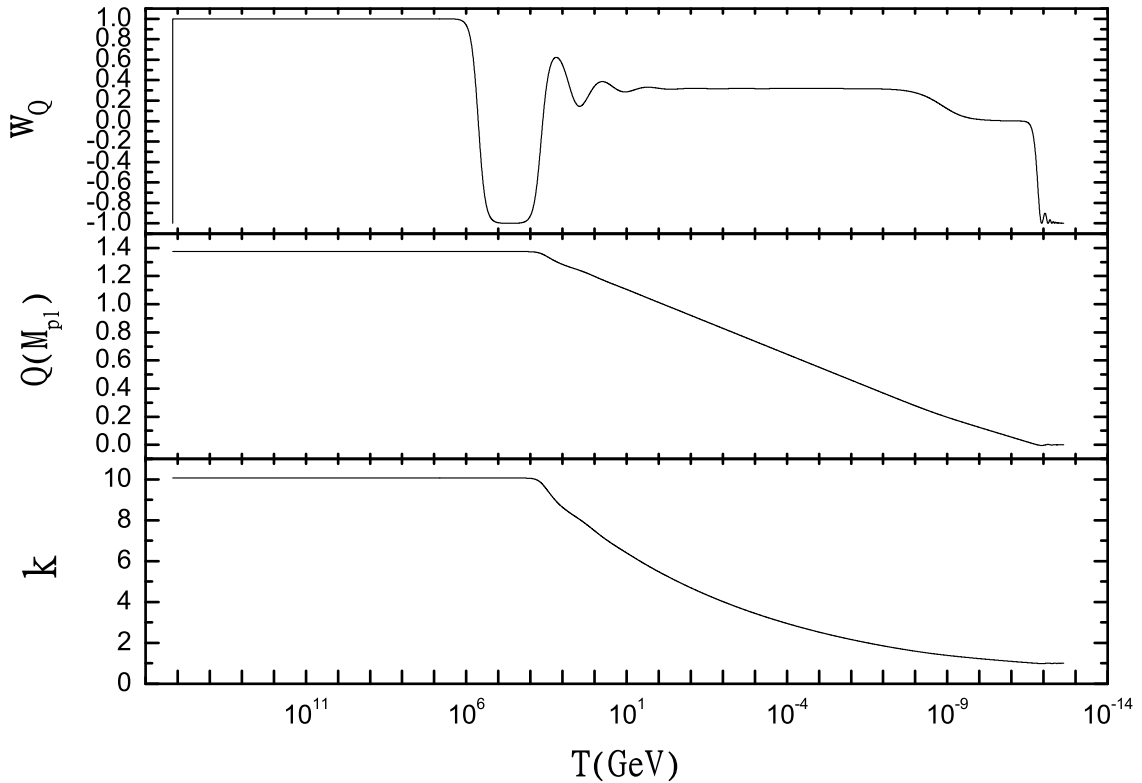}
\caption{\label{4} The evolution of $w_Q$, $Q$ and $k$ as a
function of the temperature $T$ for the double exponential
Quintessence model including the coupling with the right-handed
neutrinos and the triplet Higgs. We take $\beta=-1.68$ and
$\overline{M}_1 =3.1 \times 10^9 GeV$.}
\end{figure}
\begin{figure}
\includegraphics[scale=1.5]{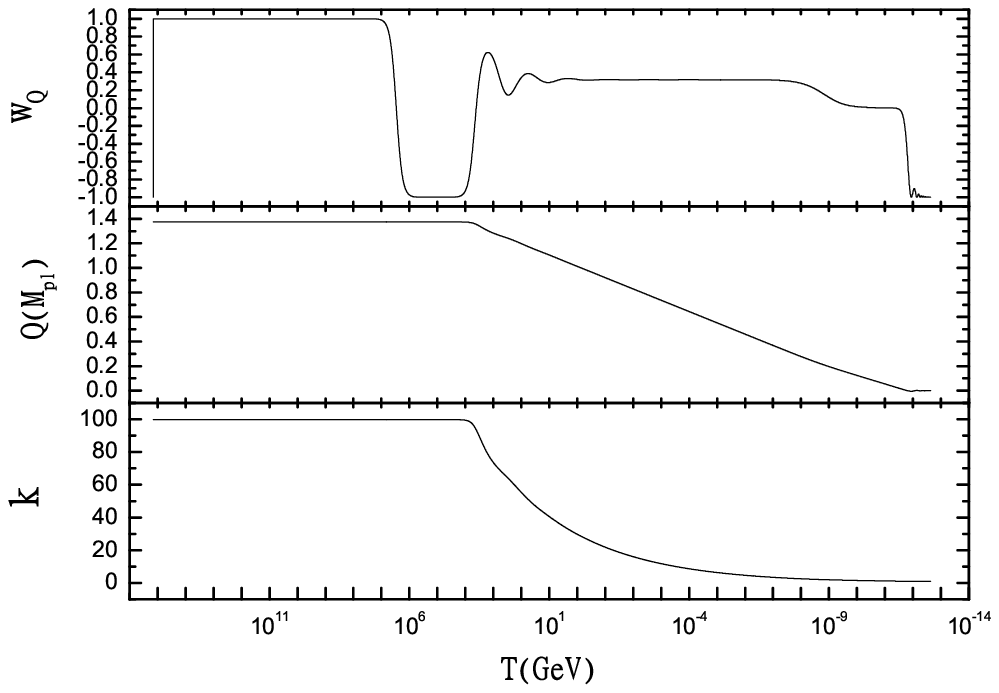}
\caption{\label{5} The evolution of $w_Q$, $Q$ and $k$ as a
function of the temperature $T$ for the double exponential
Quintessence model including the coupling with the right-handed
neutrinos and the triplet Higgs. We take $\beta=-3.35$ and
$\overline{M}_1 =3.1 \times 10^9 GeV$.}
\end{figure}
\begin{figure}
\includegraphics[scale=1.5]{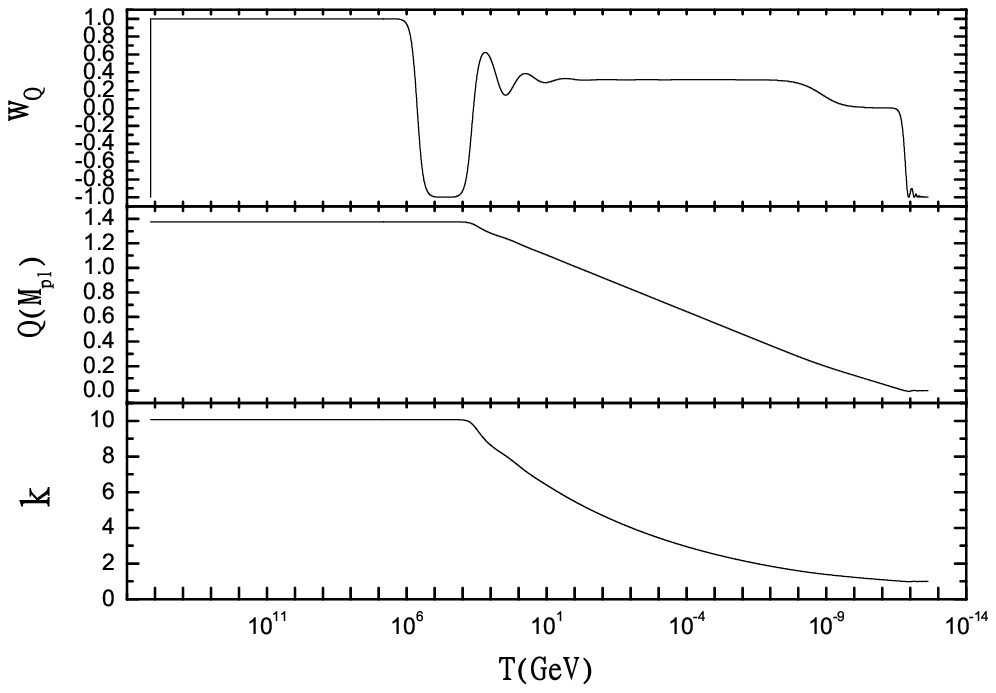}
\caption{\label{6} The evolution of $w_Q$, $Q$ and $k$ as a
function of the temperature $T$ for the double exponential
Quintessence model including the coupling with the right-handed
neutrinos and the triplet Higgs. We take $\beta=-1.68$ and
$\overline{M}_1 =2.7 \times 10^8 GeV$. }
\end{figure}
\begin{figure}
\includegraphics[scale=1.5]{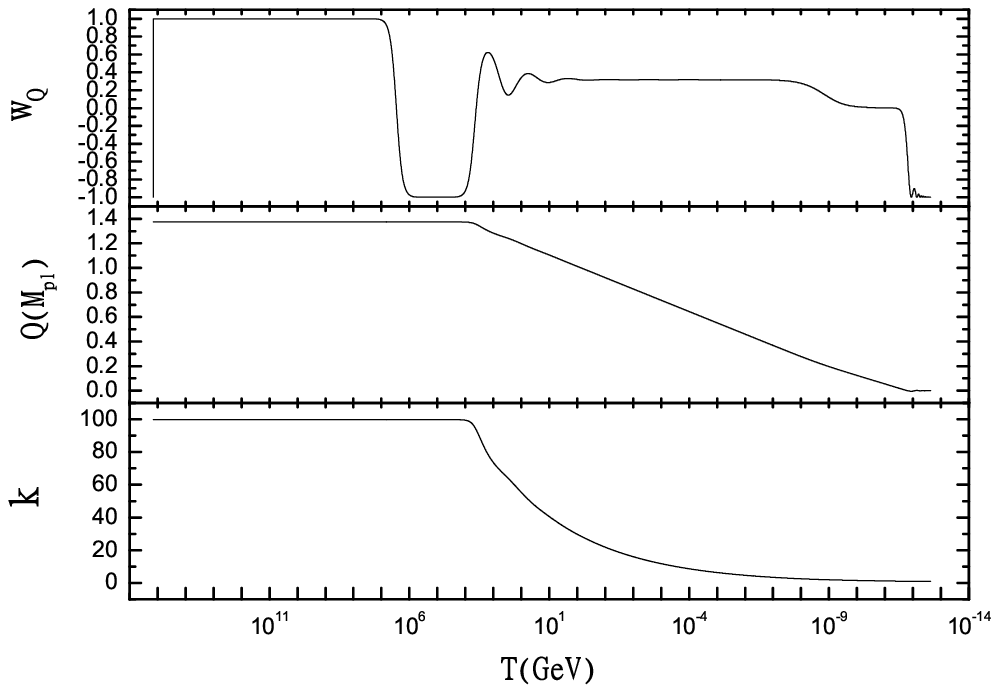}
\caption{\label{7} The evolution of $w_Q$, $Q$ and $k$ as a
function of the temperature $T$ for the double exponential
Quintessence model including the coupling with the right-handed
neutrinos and the triplet Higgs. We take $\beta=-3.35$ and
$\overline{M}_1 =2.5 \times 10^8 GeV$. }
\end{figure}

Comparing Figs. 3 with Figs. 4, 5, 6 and 7, one can see that the
interaction of the Quintessence with the right-handed neutrinos
and the triplet Higgs, does change the equation-of-state of the
Quintessence field, however, does not change the tracking
properties of this model. Furthermore, the value of the
Quintessence field $Q$ changes very little in this model until
$T\sim 10^4 GeV$ which satisfies our assumption for a constant $k$
during the period of leptogenesis.

In summary, we study the thermal leptogenesis in the scenario
where the standard model is extended to include one $SU(2)_{L}$
triplet Higgs boson, in addition to three generations of the
right-handed neutrinos. And in the model we introduce the coupling
between the Quintessence and the right-handed neutrinos, the
triplet Higgs boson, so that the light neutrino masses vary during
the evolution of the universe. Assuming that the lepton number
asymmetry is generated by the decays of the lightest right-handed
neutrino $N_{1}$, and find the thermal leptogenesis can be
characterized by four model independent parameters:
$\varepsilon_{1}, M_{1}, \bar{m}, \tilde{m}$. With the dominant
contribution of the triplet Higgs to the lepton asymmetry and the
varying neutrino masses, we find the degenerate spectrum of the
light neutrino masses and the lower reheating temperature can be
get simultaneously, by solving the Boltzmann equations
numerically.

 { \bf
Acknowledgment:} We thank Prof. Xinmin Zhang and Dr. Bo Feng for
discussions. This work is supported in part by the National
Natural Science Foundation of China under the Grand No. 90303004,
10105004.

\end{document}